\documentclass[showpacs,preprintnumbers,twocolumn]{revtex4}
\usepackage{graphicx}

\begin{document}
\title {Ferromagnetic resonant tunneling diodes as spin polarimeters and
polarizers}
\author{Francesco Giazotto}
\email{giazotto@sns.it} 
\author{Fabio Taddei$^{(1)}$} 
\email{taddei@sns.it}
\author{Rosario Fazio}
\author{Fabio Beltram}
\affiliation{NEST-INFM \& Scuola Normale Superiore, I-56126 Pisa, Italy\\
$^{(1)}$ISI Foundation, Viale Settimio Severo, 65, I-10133 Torino, Italy}

\begin{abstract}
A method  for measuring the degree of spin polarization of  magnetic materials
 based on \emph{spin}-dependent resonant tunneling is proposed. 
The device we consider is  a ballistic double-barrier resonant structure 
consisting of a ferromagnetic layer embedded between two insulating barriers. 
A simple 
procedure,  based on a detailed 
analysis of the differential conductance, allows to accurately determine the polarization of the ferromagnet. The spin-filtering character of such a system is furthermore 
addressed. We show that a  $100\%$ spin  selectivity can be achieved under appropriate conditions. 
This approach is believed to be well suited for the investigation of diluted magnetic semiconductor heterostructures. 
\end{abstract}


\maketitle

The operation of spintronic devices~\cite{Wolf} requires
the availability of efficient techniques to inject and 
detect  spin-polarized currents. In fact the search for new materials with a high degree of 
spin-polarization represents today an important technological challenge in spintronics.
A key parameter in  these devices is the spin polarization of the electric current defined as 
\begin{equation}
\mathcal P=(I_{\uparrow}-I_{\downarrow})/(I_{\uparrow}+I_{\downarrow})~, 
\end{equation}
where $I_{\sigma}$ is the contribution to the current due to $\sigma$-spin carriers \cite{Mazin}.
Structures containing ferromagnetic materials are obvious candidates as sources of spin-polarized currents.
These can be metallic~\cite{yuasa} or consist
of diluted magnetic semiconductors (DMSs)~\cite{Ohno}.
Two different electric transport methods have been devised for determining this quantity
both based on contacting the ferromagnets under investigation to a superconductor (S).
In the first one, $\mathcal P$ can be estimated by realizing a F/S tunnel junction and Zeeman-splitting 
the superconducting density of states (DOS) through the application of an external magnetic field~\cite{TM1}.
More recently the point-contact Andreev reflection (PCAR) method was introduced.
Here the suppression of Andreev reflection (AR) at F/S ballistic junctions is exploited~\cite{Ji}.
This method proves most effective for highly-transmissive F/S contacts where the AR probability is 
high~\cite{parker}, consequently its application in DMS materials is hindered by the presence of an 
unavoidable Schottky barrier at the interface.  

\begin{figure}[h!]
\begin{center}
\includegraphics[width=6.5cm]{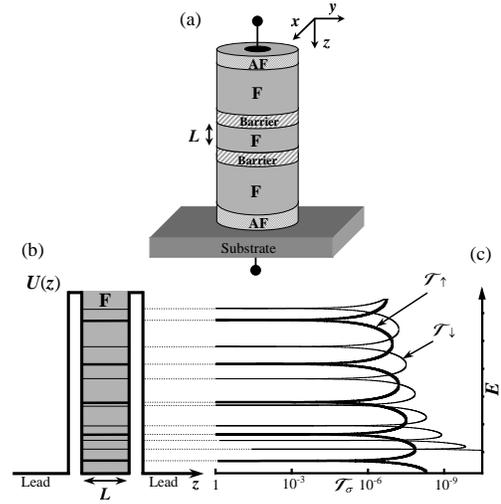}
\end{center}
\caption{(a) Scheme of a possible implementation of a ferromagnetic resonant tunneling diode. 
The magnetic layers (F), separated by insulating barriers, may  either consist of metallic 
ferromagnets or dilute magnetic semiconductors (DMSs), and AF denote antiferromagnetic layers 
that pin the direction of the electrodes magnetization (see text). (b) Schematic potential profile 
$U(z)$ of a magnetic double-barrier structure. (c) Qualitative behavior of the transmission 
probability per spin ($\mathcal{T}_{\sigma}$) of the structure. The presence of an exchange 
field in F removes the spin-degeneracy of the resonant levels thus changing their energies, 
as sketched in (b).}
\label{ARscheme}
\end{figure}

In this letter we propose an alternative route to determine
the polarization of a ferromagnet (either metallic or
DMS) based on a  double-barrier resonant structure (see Fig. 1(a)).
The operation principle of this device resides in the fact that, when the resonator is ferromagnetic,
the resonant levels of the two spin species occur at different energies whose
difference is a function of the exchange field in F (and therefore $\mathcal P$).
The advantage of this method stems from the fact that this energy difference is
little dependent on the quality of the barriers (provided their transmissivity is low enough 
as required for resonant tunneling to occur) and that no direct  coupling to a 
superconductor  is necessary, as imposed in PCAR. In addition the same device
can be operated as an extremely efficient spin filter~\cite{Ralph,rocca}.

The device consists of a ballistic ferromagnetic layer of thickness $L$ sandwiched between 
two non-magnetic barriers connected to  electrodes (in Fig. 1(b) 
the potential profile of the structure is sketched). In the following 
we refer to this structure  as to the ferromagnetic resonant tunneling diode (FRTD).
As long as phase-coherence in preserved, current through the FRTD can
flow via the resonant levels of the quantum well defined by the barriers.
Such resonances show up  as peaks in the differential conductance spectrum.
Since the resonator is ferromagnetic, the spin
degeneracy is lifted and the resonant levels relative to the two electron spin
species occur at different energies (see Fig. 1(c)).
By noting that this energy difference depends on the
polarization $\mathcal P$ of the ferromagnet, it is possible to determine 
$\mathcal P$  from the analysis of the resonance energy position.

In order to clearly explain how the method works, we consider
a single-band model as a generic example, and we describe the
ferromagnet (or DMS) with an effective exchange field 
$h_{exc}$ (Stoner model)~\cite{koenig}. 
If the resonator is decoupled from the contacts, the quantum well levels 
\begin{equation}
E_{\sigma}^{n_{\sigma}}= \mathcal{F}(k_{n_{\sigma}})-\sigma h_{exc} ~,
\label{dispersion}
\end{equation} 
(characterized by the spin-dependent index $n_{\sigma}$) are found from elementary 
quantum mechanics.
In Eq. (2), $k_{n_{\sigma}}$ is the quantized longitudinal quasiparticle wavevector, 
$\mathcal{F}(k)$ is a  generic dispersion relation, and $\sigma =\pm 1$ is the spin index, 
where the sign $+$($-$) corresponds to $\uparrow$($\downarrow$) spin channel. Note that 
$\mathcal{F}(k)$ contains a transverse contribution to the total quasiparticle energy.
For fixed $h_{exc}$ and  $L$ the requirement that  $E^{n_{\sigma}}_{\sigma} >0$ (i.e., resonant states above the Fermi energy $E_F$) determines the first resonance 
index $n_{\sigma}$. Due to presence of the exchange field, $n_{\sigma}$ can be in general 
quite different for the two spin species. By connecting the resonator to the electrodes 
through weakly-transmitting barriers such bound states acquire a finite width at the 
energies  $E^{n_{\sigma}}_{\sigma}$. As a result they manifest themselves as peaks in 
the differential conductance spectrum with a broadening dependent on barrier strength 
and temperature.

As in the PCAR method, our aim is to determine  $h_{exc}$, and therefore $\mathcal{P}$, 
assuming that the band structure parameters contained in $\mathcal{F}(k)$ are known.
Since the exchange field enters the dispersion relation additively,
one might think that it can be determined from the energy difference 
between a spin-up and a spin-down resonance relative to the same index $n_{\sigma}$.
This however is possible only if $h_{exc}$ is small with
respect to such energy difference, a situation that is typically
not fulfilled in ferromagnets and DMS.
Nevertheless, $h_{exc}$ can still be determined through an analysis of the low-temperature 
differential conductance spectrum.
One can employ the following simple procedure:
i) one identifies two successive spin-up resonances and measures their
energy difference 
from which the index $\bar{n}_\uparrow$ of the first chosen spin-up resonance 
is determined solving Eq. (\ref{dispersion});
ii) the same operation is performed on spin-down resonances in order  
to obtain $\bar{n}_\downarrow$;
iii) finally one measures the energy difference $\delta E_{\uparrow \downarrow}$ 
between one spin-up resonance of i) and one spin-down resonance of ii) so that 
the exchange field is easily obtained from
\begin{equation}
h_{exc}=\frac{1}{2}\left [ \mathcal{F} (k_{\bar{n}_{\uparrow}}) -
\mathcal{F} (k_{\bar{n}_{\downarrow}}) -  \delta E_{\uparrow \downarrow} \right ].
\end{equation}
It is worth stressing that for linear or constant dispersion curves  the proposed 
method fails. In these situations indeed the energy difference of points i) and ii) 
would be independent of the resonance index, thus preventing its determination. 
To identify the spin character of the resonances we envision the following two possibilities.
The first one is to make use of ferromagnetic leads~\cite{spin}, so that the
amplitude of the resonances will be spin-dependent (for the same reason for which, for example, the Sharvin conductance of a ferromagnetic point contact of area $\mathcal A$ and Fermi wavevector $k_{\sigma}$, $G_{Sharvin}=(e^2 /h) (\mathcal{A}\,k_{\sigma}^2 /4\pi)$, is different for the two spin species).
The second possibility relies on the application of an external  magnetic 
field which will produce a Zeeman shift of the resonances in opposite energy direction 
for the two spin species~\cite{g}.  

So far we presented an ideal situation. In order to check the actual feasibility of 
the procedure outlined above we have calculated, within a transfer-matrix approach, 
the differential conductance spectrum of a three-dimensional FRTD with rectangular 
barriers, assuming translational invariance in the plane of the junctions and large 
spin-flip length in the ferromagnet with respect to the resonator width~\cite{spinflip}.
For definiteness, we assumed a parabolic dispersion relation \cite{FRT}
and typical parameters relative to a metallic ferromagnet \cite{calcolo}.
The resonator width was chosen in order to show a few  resonances per spin
within a bias voltage range of one tenth of  $E_F$. 
Barrier transmissivity ($\approx 10^{-2}$) was taken such that the resonances width is small with respect to their relative separation, and a finite temperature was also 
taken into account in order to make the simulation more realistic.
Having computed the differential conductance spectrum, as if obtained directly from a
measurement, we have identified the peaks positions and spin character.
By applying our procedure we have determined $h_{exc}$ with a negligible error 
with respect to the nominal value. This proves the effectiveness of this method.
Furthermore, it is remarkable that the method works properly  even if the 
dispersion curve for bias voltages in the range considered is not far from linearity.
We wish to remark that all the above simulations are just an exemplification
of how the method works and that some experiments might need to be analyzed 
employing, for example, multi-band models.
In this last case the method is still valid even though
the analysis becomes  more cumbersome.

\begin{figure}[h!]
\begin{center}
\includegraphics[width=5cm]{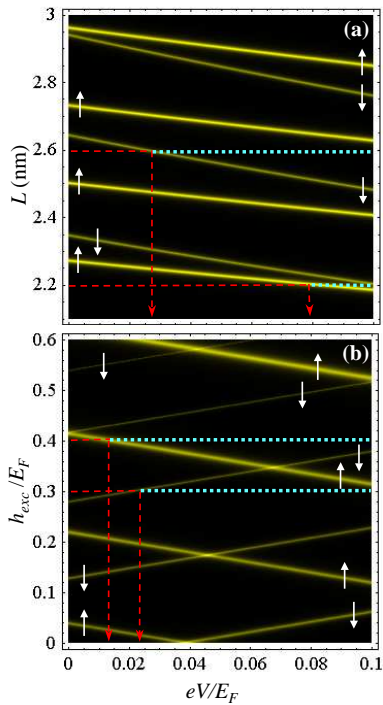}
\end{center}
\caption{ Contour plot of the zero-temperature differential conductance spectrum 
(arbitrary units) versus bias voltage and resonator width (a), and versus  exchange 
field (b). White arrows indicate the spin character of the resonances and the blue 
dashed-lines show  examples of possible bias ranges in which a $100\%$ spin-polarized 
current is achieved for chosen $L$ (a), or fixed $h_{exc}$ (b).
In these simulations we used $E_F =5.7\,eV$, barrier transmissivity of $10^{-1}$, 
$h_{exc}=1.425\,eV$ (a) and $L=3$ nm (b). The color scale ranges from black to yellow 
indicating a variation of conductance from zero to maximum, and yellow brightness 
is proportional to the resonances amplitude.
}
\label{ARscheme1}
\end{figure}

Finally we  discuss  the spin-filtering action of the proposed structure.
For bias voltages  such that only the \emph{first} resonance contributes to 
transport, i. e., for voltages which correspond to the first step in the 
current-voltage characteristic~\cite{Ralph}, the system behaves as an ideal spin filter, 
providing fully-polarized ($100\%$) currents even in the absence of a \emph{half-metallic} (i.e., with $\mathcal P =1$) 
resonator. The spin polarization of the current  can be chosen either by varying 
the diode geometry (the resonator width  $L$) or by selecting an appropriate 
material (with a given $h_{exc}$).
In both situations, the application of an external magnetic field offers an additional 
way of tuning the spin-filtering character of the structure.
To elucidate this property  we show in Fig. 2(a) a  contour plot of the zero-temperature 
differential conductance as a function of resonator width   and bias voltage. In this plot we 
assumed that the  FRTD electrodes are of the same magnetic material as the resonator. 
The figure clearly shows that the spin character of the first resonance (indicated by 
$\uparrow ,\downarrow$ white arrows) can be changed upon variation of $L$. For example 
by choosing $L=2.2$ nm the first resonance occurs at $eV=0.08 \,E_F$ and is spin-up in 
character, while choosing $L=2.6$ nm a spin-down resonance occurs at $eV=0.027 \,E_F$ 
(see red dashed-lines).
A similar dependence is observed in Fig. 2(b) where the differential conductance is plotted  
versus the exchange field for fixed resonator width ($L=3$ nm).

We conclude by elaborating on the actual realization of the FRTD as a spin polarizer. 
Figure 1(a) shows a FRTD where the magnetization direction of the two ferromagnetic 
electrodes (F) is pinned through their contact with antiferromagnetic layers 
(AF), and the magnetization of the resonator can be changed from parallel to antiparallel 
direction upon the application of weak magnetic fields. The device could be engineered 
through the sequence of  metallic ( for example Ni$_x$Fe$_{1-x}$ alloy)  or III-V DMSs (typically 
Ga$_{1-x}$Mn$_{x}$As) stacked layers, and taking advantage of the $x$-dependence of their 
exchange field in order to tune the spin-filter.    

In summary, we have proposed a general method which exploits a FRTD for determining 
the degree of polarization of a ferromagnet. The spin-filtering character of such 
structure was additionally addressed and ideal spin selectivity proven.
The proposed procedure seems to be more appropriate than PCAR for the
case of DMSs, where the formation of
a transmissive contact with the superconductor is expected to be difficult due
to the presence of an unavoidable Schottky barrier at the interface with the
metal.
On the contrary,  the technology for the fabrication of magnetic resonant tunneling diodes, both metallic~\cite{Ralph,RT} and consisting of DMSs~\cite{RT2}, is
nowadays already available.

The authors gratefully acknowledge S. Sanvito for useful discussions. 
This work was supported by INFM under the PAIS project TIN and 
European Community through grant (RTN2-2001-00440).



\begin{references}
\bibitem{spin}
	Due to small value of the involved coercitive fields, a small magnetic field 
	is sufficient to make sure that the magnetization of all layers are parallely-aligned.
\bibitem{spinflip}
	The presence of spin-flip scattering does not alter the energy position 
	of the resonances. When using ferromagnetic electrodes it simply alters 
	the relative amplitude of  resonances belonging to different spin species.
\bibitem{calcolo}
	In particular we employed $E_F =5.7\,eV$ and $h_{exc}=1.425\,eV$, 
	corresponding to $\mathcal P =25 \%$.
\end{references}
\end{document}